\documentclass[compactaffiliation]{Interspeech}
\usepackage{multirow}
\usepackage{pifont}
\usepackage{fontawesome} 

\interspeechcameraready

\title{SOVA-Bench: Benchmarking the Speech Conversation Ability for LLM-based Voice Assistant}

\author[affiliation={1},equalcontribution]{Yixuan}{Hou}
\author[affiliation={1,2},equalcontribution]{Heyang}{Liu}
\author[affiliation={1,2}]{Yuhao}{Wang}
\author[affiliation={3}]{Ziyang}{Cheng}
\author[affiliation={2}]{Ronghua}{Wu}
\author[affiliation={2},corresponding]{Qunshan}{Gu}
\author[affiliation={1}]{Yanfeng}{Wang}
\author[affiliation={1},corresponding]{Yu}{Wang}

\affiliation{School of Artificial Intelligence}{Shanghai Jiao Tong University}{China}
\affiliation{}{Ant Group}{China}
\affiliation{School of Cyber Science and Engineering}{Wuhan University}{China}

\email{\{yixuan303a,liuheyang,colane,wangyanfeng622,yuwangsjtu\}@sjtu.edu.cn,\\\{r.wu,guqunshan.gqs\}@antgroup.com,icelookgoose@gmail.com}

\usepackage{comment}

\begin{document}

\maketitle
\renewcommand{\thefootnote}{\fnsymbol{footnote}}
\footnotetext[1]{Corresponding Authors}
\renewcommand{\thefootnote}{\arabic{footnote}}

\begin{abstract}
    
        Thanks to the steady progress of large language models (LLMs), speech encoding algorithms and vocoder structure, recent advancements have enabled generating speech response directly from a user instruction. However, benchmarking the generated speech quality has been a neglected but critical issue, considering the shift from the pursuit of semantic accuracy to vivid and spontaneous speech flow. Previous evaluation focused on the speech-understanding ability, lacking a quantification of acoustic quality. In this paper, we propose \textbf{S}peech c\textbf{O}nversational \textbf{V}oice \textbf{A}ssistant \textbf{Bench}mark (\textbf{SOVA-Bench}), providing a comprehension comparison of the general knowledge, speech recognition and understanding, along with both semantic and acoustic generative ability between available speech LLMs. To the best of our knowledge, SOVA-Bench is one of the most systematic evaluation frameworks for speech LLMs, inspiring the direction of voice interaction systems.
    
\end{abstract}

\section{Introduction}

Communication through speech outperforms text interaction for its convenience and efficiency. Recent advancements in large language models (LLMs) have led to remarkable breakthroughs in speech LLMs and voice assistants. Starting from an LLM trained on tremendous textual corpora, these models perceive speech using an encoder with adaptors, generating the capability to understand speech flow through supervised fine-tuning, even producing speech response directly with a vocoder. The most promising application of generative speech LLMs lies in the voice assistant. It is proposed to respond to the user's vocal instructions with vivid and spontaneous speech, enabling accurate and efficient interactions. One of most powerful LLM, GPT-4o \cite{hurst2024gpt}, has already supported speech modality interaction, and then this ability has spread to open-sourced models, such as Mini-Omni \cite{xie2024mini}, LLaMA-Omni \cite{fang2024llama} and Moshi \cite{defossez2024moshi}, showing the promising path for speech-central human-machine interaction.


The lack of an evaluation system is one of the main factors limiting the development of generative speech LLMs. Previous work has predominantly adopted benchmarks for the model's understanding capabilities. Dynamic SUPERB provides a platform to compare various models on plenty of speech-processing tasks \cite{huang2024dynamic, huang2024dynamic2}. AudioBench and AIR-Bench focus on the performance of understanding various audio signals including human speech and natural sound \cite{wang2024audiobench, yang2024air}. VoiceBench measures voice assistants for their capability of general knowledge, instruction following, robustness and safety alignment \cite{chen2024voicebench}. However, the above benchmarks either only measure the model's ability to produce text responses or transform the speech response via an automatic speech recognition (ASR) system. They assess understanding or text-based transformations, but fall short in evaluating the quality of the voice output itself—an essential factor for voice assistants and other interactive systems. 


\begin{figure}[t!]
    \centering
    \includegraphics[width=1\linewidth]{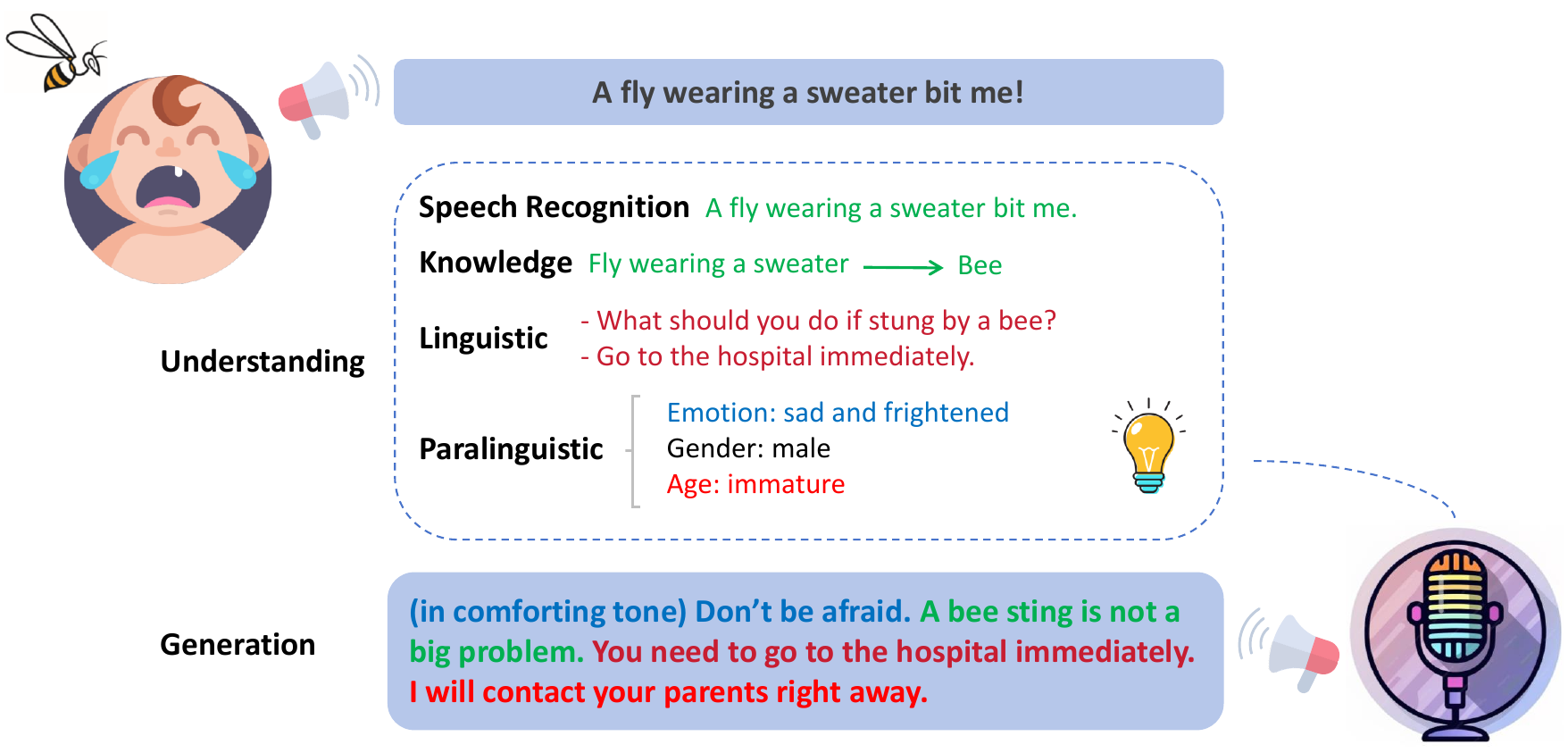}
    \caption{Properties for an ideal voice assistant. The color of the generated response corresponds to distinct aspects of the user's query. For example, if the model receives emotions of sadness and fear, a comforting tone should usually be chosen.}
    \label{fig:properties_va}
    \vspace{-0.6cm}
\end{figure}


To address this gap, we introduce an exhaustive benchmark, \textbf{S}peech c\textbf{O}versational \textbf{V}oice \textbf{A}ssistant \textbf{Bench}mark (SOVA-Bench), to quantify the performance of generative speech LLMs, especially voice assistants. As shown in Figure~\ref{fig:properties_va}, an ideal voice assistant should respond to user queries using preserved general knowledge based on instruction content, linguistic, and paralinguistic information. SOVA-Bench evaluates the models with distinct subsets. It ingests raw audio from multiple datasets and presents in the format of question-answering (QA) and instruction querying. By encompassing both understanding and generation, including measures for tone, emotion, and other paralinguistic cues, this method provides a more comprehensive evaluation framework that mirrors real-world communication requirements. Ultimately, it addresses a significant gap in the field, paving the way for the development of advanced generative speech models that deliver not only accurate information but also a reliable, engaging, and human-like auditory experience. Our main contributions are as follows:


\begin{table*}[htbp!]
\caption{Comparison with other available speech LLM evaluation benchmarks.}
\vspace{-0.2cm}
\resizebox{1\linewidth}{!}{
\begin{tabular}{l|c|c|c|c|c|c|c}
\toprule
 \multirow{2}*{\bf Benchmark}& \multirow{2}*{\bf Num of Tasks} & \multirow{2}*{\bf Knowledge} & \multirow{2}*{\bf Recognition} & \multicolumn{2}{c|}{\bf Understanding} & \multirow{2}*{\bf Generation} & \multirow{2}*{\bf Evaluation Subject} \\ 
 &  & &  & \multicolumn{1}{c|}{Linguistic} & \multicolumn{1}{c|}{Paralinguistic} & &\\ \midrule

 Dynamic SUPERB Phase-2 & 180 & \ding{56} & \ding{52} & \ding{52} & \ding{52} & \ding{56} & Universal Speech Models \\ 

 AudioBench & 8 & \ding{56} & \ding{52} & \ding{52} & \ding{52} & \ding{56} &  AudioLLMs  \\

 AIR-Bench & 20 & \ding{56} & \ding{52} & \ding{52} & \ding{52} & \ding{56} &  AudioLLMs  \\

 VoiceBench & 6 & \ding{52} & \ding{56} & \ding{52} & \ding{56} & \ding{56} & LLM-based Voice Assistants  \\

 OpenAudioBench & 3 & \ding{52} & \ding{56} & \ding{52} & \ding{56} & \ding{56} & Audio-focused LLMs  \\

 \bf SOVA-Bench (Ours) & 8 & \ding{52} & \ding{52} & \ding{52} & \ding{52} & \ding{52} & Generative Speech LLMs  \\

\bottomrule
\end{tabular}
}
\label{tab:related_benchmark}
\vspace{-0.5cm} 
\end{table*}









\begin{itemize}
    \item We propose SOVA-Bench, an evaluation system for generative speech LLMs and voice assistants, quantifying the performance of general knowledge, as well as the ability to recognize, understand and generate speech flow.
    \item SOVA-Bench compares multiple generative speech models under the same evaluation framework. Based on the comparison experiments, we identify the limitations of existing voice interaction models.

    \item SOVA-Bench provides a standard for the evaluation of voice interactive systems and facilitates subsequent researchers to compare model performance.
    
\end{itemize}

\section{Related Works}

\subsection{Speech LLM}

Speech LLM extends the understanding capability to speech flow, performing modality alignment between speech and text via an encoder with adaptors. Earlier work focused on generating textual responses from speech instructions, optionally adopting a TTS module for a cascaded speech-in-speech-out paradigm, such as SALMONN \cite{tangsalmonn}, SpeechGPT \cite{zhang2023speechgpt} and Qwen2-audio \cite{chu2024qwen2}. These models preserve the ability to understand speech signals while show a significant difference from human-like real-time conversations and voice assistants. The real-time API introduced by GPT-4o supports end-to-end speech generation without an explicit TTS module. It can produce text and speech responses at the same time, enabling fluent and efficient communication with users. By utilizing speech vocoder methods, several open-sourced models have also achieved speech response from discrete acoustic tokens. Mini-Omni \cite{xie2024mini} adopted SNAC \cite{siuzdak2024snac} to quantify the model's response speech during training with a transformer-based streaming audio decoder to covert the predicted embedding to speech flow. LLaMA-Omni \cite{fang2024llama} alternatively discrete continuous speech to units using pre-trained HuBERT \cite{hsu2021hubert}, and a HiFi-GAN vocoder \cite{kong2020hifi, polyak2021speech} for unit-to-speech. Moshi \cite{defossez2024moshi} proposed a novel speech encoder, Mimi, incorporating both semantic and acoustic speech embeddings for information extraction and generative speech wave construction. Other speech LLMs that enable direct speech response include Westlake-Omni \cite{WestlakeOmni}, Freeze-Omni \cite{wang2024freeze}, SyncLLM \cite{veluri2024beyond}, OmniFlatten \cite{zhang2024omniflatten}, etc. These models mainly follow a similar schema - generate discrete audio tokens using LLM before converting to speech signals. The generative speech LLMs support real-time communications, showing great potential for voice assistants.

\subsection{Speech LLM evaluation}

The evaluation of speech models is a contentious issue. For specialized models, the comparison between reference and output is widely adopted, such as automatic speech recognition (ASR) and speech emotion recognition (SER) models. For speech LLMs with multiple capabilities, comprehensive evaluation benchmarks have been proposed to systematically assess a model's performance, as shown in Table~\ref{tab:related_benchmark}. Dynamic SUPERB is a universal speech processing benchmark consisting of 33 tasks to evaluate speech models in a zero-shot fashion, and then extend to 180 tasks in the 2nd phase \cite{huang2024dynamic, huang2024dynamic2}. AudioBench presents a set of measurements targeting at linguistic and paralinguistic speech understanding, and audio scene recognition \cite{wang2024audiobench}. AIR-Bench encompasses a series of tasks to measure the performance of understanding human speech, natural sound, and music \cite{yang2024air}. It preserves two dimensions - the foundation benchmark for specialized ability and the chat benchmark for general performance. These two benchmarks are originally designed for Audio LLMs. VoiceBench aims at a multifaceted evaluation for LLM-based voice assistants \cite{chen2024voicebench}. It distills proper instances to access the model's capability of general knowledge, instruction following and safety alignment, along with the robustness of speaker, environment and content variation. OpenAudioBench is designed to assess the capabilities of multimodal and audio-focused language models, consisting of subsets of logical reasoning, general knowledge and open-ended questions \cite{li2025baichuan}. The above-mentioned benchmarks show the same limitation, for they all concentrate on the understanding capability while ignoring the generative performance.

\section{SOVA-Bench}

As shown in Table~\ref{tab:dataset_statistics}, SOVA-Bench aims to comprehensively evaluate and compare speech LLMs' abilities to perceive instructions and generate responses both in speech formats, therefore contributing to the evolution of voice assistants.

\subsection{Voice Assistant Properties}

Voice assistants are intelligent systems designed to enhance human-computer interaction through natural speech processing, understanding and generation. An ideal voice assistant shall preserve such properties:

\begin{itemize}
    \item \textbf{General Knowledge}: Voice assistants accumulate massive and diverse general knowledge far beyond the level of normal human mastery. When the user proposes an inquiry, the system responds accurately with high credit and confidence. 

    \item \textbf{Speech Recognition}: Voice assistants enable speech recognition, which is not only a common task benefiting user experience but also contributes to speech understanding.

    \item \textbf{Speech Understanding}: Voice assistants should understand speech in both linguistic and paralinguistic dimensions. For the former part, the system distills the semantic information and responds accordingly. For the paralinguistic part, it should generate messages such as the user's identification and emotion, therefore producing personalized responses.

    \item \textbf{Speech Generation}: Voice assistants generate vivid and spontaneous speech responses when being evoked. The generated speech should preserve high accuracy, fluency, and clarity, and be consistent with its text transcription. The speech response quality should be evaluated from the perspective of both semantic and acoustic.

\end{itemize}

\begin{table*}[htbp]
\caption{Statistics for SOVA-Bench}
\vspace{-0.2cm}
\resizebox{1\linewidth}{!}{
\begin{tabular}{l|c|c|c|c|c|c|c|c|c}
\toprule
 \multirow{2}*{\bf Dimension}& \multirow{2}*{\bf Task} & \multirow{2}*{\bf Source Data} & \multirow{2}*{\bf Synthesized} & \multirow{2}*{\bf Format} & \multicolumn{2}{c|}{\bf Input Modality} & \multirow{2}*{\bf Num of Instances} & \multicolumn{2}{c}{\bf Evaluation}\\ 
 
 & & & & & \multicolumn{1}{c|}{Text} & \multicolumn{1}{c|}{Speech} &  & Method & Metrics \\ \midrule

Knowledge & Knowledge QA & TriviaQA & \ding{52} & Open-domain & \ding{56} & \ding{52} & 6840 & GPTEval & Accuracy \\ \midrule

Recognition & Speech Recognition & LibriSpeech & \ding{56} &  & \ding{52} & \ding{52} &  & Normal & WER \\ \midrule

\multirow{2}*{Linguistic} & \multirow{2}*{Spoken QA} & LibriSQA & \ding{56} & Multi-choice (4) & \ding{52} & \ding{52} & 2619 & Normal & Accuracy \\ 

 & & Spoken SQuAD & \ding{52} & Open-domain & \ding{52} & \ding{52} & 3243 & GPTEval & Accuracy \\ \midrule

 \multirow{3}*{Paralinguistic} & Emotion Recognition & IEMOCAP & \ding{56} & Multi-choice (5) & \ding{52} & \ding{52} & 5819 & GPTEval & Accuracy \\ 

 & Gender Recognition & Common Voice & \ding{56} & Multi choice (2) & \ding{52} & \ding{52} & 2469 & GPTEval & Accuracy \\

 & Age Recognition & Common Voice & \ding{56} & Multi-choice (5) & \ding{52} & \ding{52} & 2511 & GPTEval & Accuracy \\ \midrule

 \multirow{3}*{Generation} & Consistency & Alpaca & \ding{52} & Open-domain & \ding{56} & \ding{52} & 3745 & Normal & WER \\ 

 & Semantic & Alpaca & \ding{52} & Open-domain & \ding{56} & \ding{52} & 3745 & GPTEval & GPTScore \\

  & Acoustic & Alpaca & \ding{52} & Open-domain & \ding{56} & \ding{52} & 3745 & UTMOSv2 & Score \\
 
\bottomrule

\end{tabular}
}
\label{tab:dataset_statistics}
\vspace{-0.4cm} 
\end{table*}

\subsection{Source Dataset}

\textbf{General knowledge:} To evaluate the general knowledge performance of speech LLMs, SOVA-Bench incorporates TriviaQA, an open-domain question-answering (QA) dataset with multiple reference resources \cite{joshi2017triviaqa}. We utilize the unfiltered dev set and remove the reference to form a knowledge QA dataset.

\noindent \textbf{Speech recognition:} For ASR performance, we evaluate the model's performance with LibriSpeech \cite{panayotov2015librispeech}. All of the four evaluation sets are tested for recognition accuracy.

\noindent \textbf{Speech understanding:} To measure the semantic understanding of speech information, SOVA-Bench consists of two formats of spoken QA - LibriSQA for multi-choice form and Spoken SQuAD for open-ended test \cite{zhao2024librisqa, lee18d_interspeech}. LibriSQA is developed based on LibriSpeech with the help of ChatGPT, sharing the same speech content. The comparison between the ASR and SQA performance can reveal the inner process of models to perceive and understand speech semantic information. In our benchmark, we include the second part of LibriSQA to form a simpler spoken language understanding task. Spoken SQuAD is an open-domain QA dataset, while the right answer is included in the speech transcription. We preserve a single question for each valid sentence. For those with multiple answers, we preserve the most complete one if they are in an inclusion relationship, otherwise all answers are considered correct. For paralinguistic understanding, SOVA-Bench includes emotion recognition, gender recognition and age prediction. These tasks cover the necessary user information required by voice assistants. SOVA-Bench incorporates IEMOCAP for emotion recognition \cite{busso2008iemocap}. The original source includes 10 emotions, and we delete neutral kinds and merge indistinguishable categories to form a 5-choice selection task: angry, happy, fearful, frustrated and sad, excited and surprised. For gender and age recognition, we use the latest version of Common Voice \cite{ardila2020common}. Only those with speaker information annotations are preserved. 

\noindent \textbf{Speech generation:} SOVA-Bench is developed for voice assistants, following an instruction-inquiry scenery in the generation performance evaluation. We use a cleaned version of Alpaca \cite{taori2023stanford}. We select a small portion and ensure the preserved instances without additional input. This part includes user questions and instructions. The speech generation performance is evaluated across 3 dimensions: the consistency between text and speech response, and the semantic and acoustic quality. The consistency is measured by using text response as transcription and comparing with speech output. The semantic performance is evaluated with GPT, and we implement one advanced MOS predictor, UTMOSv2 \cite{baba2024t05}, to quantify the acoustic score.


\begin{figure}[htbp]
    \centering
    \includegraphics[width=0.95\linewidth]{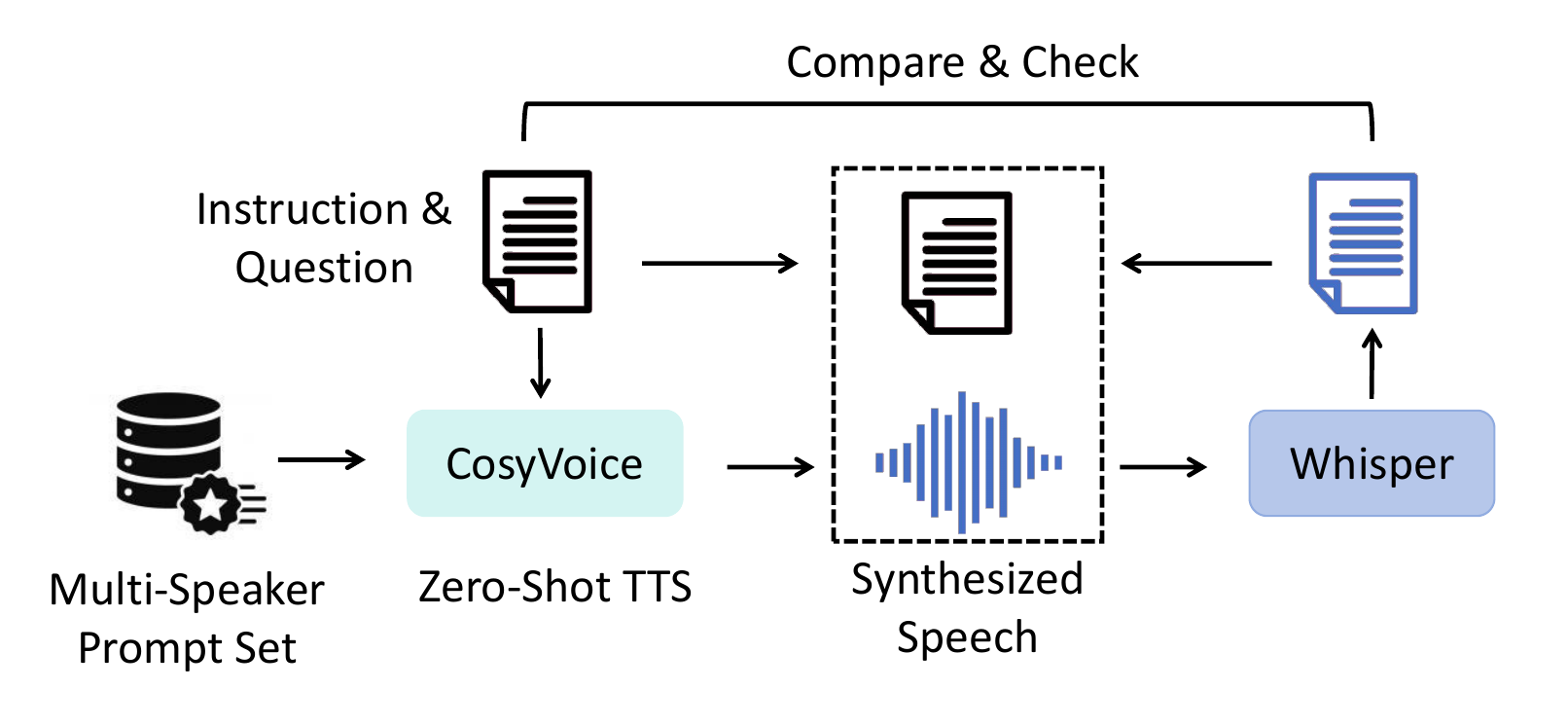}
    \caption{Creation pipeline for dataset without speech modality. A multi-speaker prompt set is adopted for zero-shot TTS.}
    \label{fig:speech_ann}
    \vspace{-0.4cm}
\end{figure}

It should be noted that neither TriviaQA nor Alpaca has speech audio. We implement an advanced TTS model, CosyVoice \cite{du2024cosyvoice}, to synthesize speech audio in a zero-shot fashion. The speech prompt is derived from Common Voice, and we select instances with speaker labels and ensure the gender-ratio to 1:1. The synthesized speech is processed by Whisper-large-v3 \cite{radford2023robust} to compare with the original transcription, and only those with high accuracy are preserved. As illustrated in Figure~\ref{fig:speech_ann}, this synthesized method ensures the diversity of speakers and acoustic environments, thereby more robustly evaluating the performance.



\section{Evaluation Experiment}

\subsection{Evaluated Models}

We have tested various widely used voice assistants on SOVA-bench, as shown in Table~\ref{tab:tested_models}, including: Mini-Omni \cite{xie2024mini}, Mini-Omni2 \cite{mini2}, LLaMA-Omni \cite{fang2024llama}, Freeze-Omni \cite{wang2024freeze}, Moshi \cite{defossez2024moshi}, and GLM-4-Voice \cite{zeng2024glm}. Additionally, we developed a cascade model for comparative evaluation, which combines Whisper-large-v3 as the ASR model and GPT-4o-mini as the LLM. Among these models, only LLaMA-Omni supports synchronous inputs of both text and speech modalities, while Mini-Omni and Mini-Omni2 support either text or speech at a time.

\begin{table}[htbp!]
\vspace{-0.1cm}
\caption{The summarization of the tested speech LLMs. The `Modality' column represents the input modalities, where `T' and `S' represent text and speech respectively.}
\vspace{-0.2cm}
\resizebox{1\linewidth}{!}{
    \begin{tabular}{l|c|c|c}
        \toprule
        \multirow{2}*{\bf Speech LLM} & \multicolumn{2}{c|}{\bf Architecture} & \multirow{2}*{\bf Modalities} \\ \cline{2-3}
         & \multicolumn{1}{c|}{Speech Encoder} & \multicolumn{1}{c|}{Base LLM} & \\ \midrule
        
         Mini-Omni & Whisper-small & Qwen2-0.5B & $T \lor S$ \\
        
         Mini-Omni2 & Whisper-small & Qwen2-0.5B & $T \lor S$ \\
        
         Moshi & Mimi & Helium-7B & S \\
        
         LLaMA-Omni & Whisper-large-v3 & LLaMA-3.1-8B-Instruct & $T \land S$ \\
        
         Freeze-Omni & CNN + Transformer & Qwen2-7B-Instruct & S \\
        
         GLM-4-Voice & VQ + Whisper & GLM-4-9B & S \\
        
         Cascade & Whisper-large-v3 & GPT-4o-mini & S \\
        
        \bottomrule
    \end{tabular}
}
\label{tab:tested_models}
\vspace{-0.3cm} 
\end{table}

It is important to note that for Moshi and GLM-4-Voice, we randomly sampled 500 instances from each subset. Since Moshi typically begins a conversation with an introductory statement (e.g. "Hey, how is everything going?"), we include 2.5 seconds of blank audio before each input to avoid confusion.

\begin{table*}[htbp]
\caption{Evaluation performance for speech LLMs on SOVA-Bench}
\vspace{-0.2cm}
\resizebox{1\linewidth}{!}{
\begin{tabular}{lcccccccccccc}
\toprule
 \multirow{2}*{\bf Model}& \multirow{2}*{\bf LLM size} & \multirow{2}*{\bf Knowledge $\uparrow$ } & \multicolumn{2}{c}{\bf Recognition $\downarrow$} & \multicolumn{5}{c}{\bf Understanding $\uparrow$ } & \multicolumn{3}{c}{\bf Generation $\uparrow$ } \\ \cmidrule(r){4-5}  \cmidrule(r){6-10} \cmidrule(r){11-13}
 
  &  &  & dev\_clean/other & test\_clean/other & LibriSQA & Spoken SQuAD & Emotion & Age & Gender & Consistency & Semantic & Acoustic \\ \midrule

 Mini-Omni & 0.5B & 2.42 & 5.18 / 12.03 & 8.68 / 12.75 & - & - & 28.53 & 11.79 & 27.22 & 10.04 & 1.44 & 3.62 \\

 Mini-Omni2 & 0.5B & 0.37 & 7.54 / 11.58 & 3.99 / 13.44 & - & - & 24.52 & 16.05 & 16.73 & 25.92 & 1.68 & \textbf{3.66} \\

 Moshi & 7B & 4.00 & - & - & - & - & - & - & - & 5.94 & 1.57 & 2.43 \\ 

 LLaMA-Omni & 8B & 24.91 & $>$100 & $>$100 & \textbf{50.67} & \textbf{43.07} & 37.74 & 19.43 & \textbf{78.41} & \textbf{4.71} & \textbf{3.28} & 3.61 \\


 Freeze-Omni & 7B & \textbf{27.27} & \textbf{3.29} / \textbf{7.40} & 3.24 / 7.68 & - & - & \textbf{38.31} & 17.80 & 39.13 & 10.96 & 3.20 & 3.46 \\

GLM-4-Voice & 7B & 25.58 & - & \textbf{2.82} / \textbf{7.66} & - & - & 30.16 & \textbf{37.84} & 69.40 & 10.02 & 3.25 & 3.09 \\

 \midrule
 Cascade & - & \textbf{58.98} & \textbf{1.92} / \textbf{3.77} & \textbf{1.81} / \textbf{3.62} & \textbf{81.11} & \textbf{93.05} & - & - & - & - & \textbf{3.88} & - \\
 
\bottomrule

\end{tabular}
}
\label{tab:results}
\vspace{-0.3cm} 
\end{table*}

\begin{figure}[htbp]
    \centering
    \includegraphics[width=0.98\linewidth]{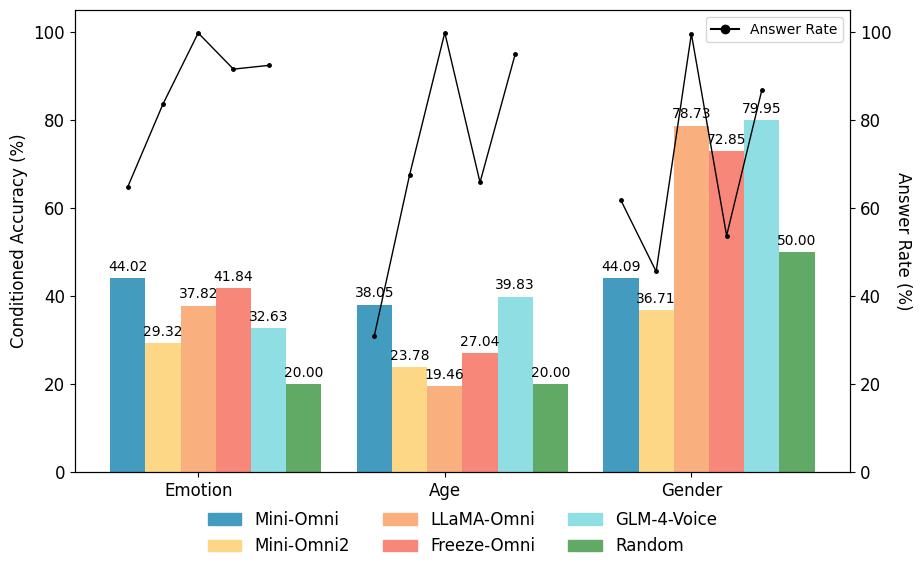}
    \caption{Performance for paralinguistic understanding.}
    \label{fig:accuracy}
    \vspace{-0.4cm}
\end{figure}





 



 

\subsection{Main Results}



The overall performance of tested speech LLMs is presented in Table~\ref{tab:results}. We observe considerable variation in performance, with no single model consistently outperforming across all tasks.

Speech models with larger LLM parameters (7B, 8B) usually outperform smaller models (0.5B) on knowledge, speech understanding and the semantic quality of generated responses. These tasks largely depend on the capabilities preserved from its LLM backbone. The cascade system preserves 58.98\% accuracy on knowledge QA, while the best speech LLM, Freeze-Omni, achieves 27.27\%. LLaMA-Omni is the only model supporting the simultaneous input of both text and speech, but exhibits a performance deficit of 30\% and 50\% in both spoken QA formats compared to the cascade method. A strong performance deterioration is observed when extending to speech modality. Moshi achieves an inferior score, as its streaming modeling results in frequent interruptions and irrelevant responses. 

Most speech LLMs demonstrate comparable performance in speech recognition. The results of Freeze-Omni and GLM-4-Voice are borrowed from their papers, as the intermediate model and specific prompts are not available. LLaMA-Omni could not perform speech recognition, for it does not involve pre-training on the ASR task, resulting in irrelevant responses given ASR instructions. Influenced by the capability of instruction following, the paralinguistic part of speech understanding tasks is also affected. Besides the overall accuracy, as shown in Figure~\ref{fig:accuracy}, we simultaneously evaluated the model's answer rate (AR) and conditioned accuracy (CA), with the latter representing the proportion of correct responses among the answered samples. LLaMA-Omni consistently exhibits high performance in instruction following; however, the best model in terms of conditioned accuracy is not the same across the three tasks. Most speech LLMs lack sufficient capability to capture specific paralinguistic information. LLaMA-Omni, Freeze-Omni, and GLM-4-Voice demonstrated clearly above-chance accuracy in gender prediction, while Mini-Omni and GLM-4-Voice showed notable performance in age prediction. Emotion recognition presents a special case for evaluation, as it is also significantly influenced by semantic content. All models demonstrate positive performance on the emotion recognition task, with Mini-Omni correctly identifying nearly half of the emotional labels.

Regarding the quality of generated speech, Moshi and LLaMA-Omni exhibit strong speech-text consistency. In contrast, Mini-omni2 demonstrates the poorest performance, primarily due to premature truncation occurring during the speech generation process. LLaMA-Omni and GLM-4-Voice exhibit the highest performance in terms of the semantic quality of generated responses. Theoretically, a larger parameter LLM backbone is conducive to generating higher-quality responses, while Moshi once again received a lower score due to an excessive amount of irrelevant output. As for the acoustic quality, Mini-Omni2 achieves the highest score, a little higher than Mini-Omni and LLaMA-Omni. The acoustic performance is affected by the speech token modeling method.

\subsection{Limitations for current speech LLMs}

Based on the experimental results from SOVA-Bench, we summarize the current limitations of speech LLMs as follows:

\begin{itemize}
    \item Limited input modalities: Only LLAMA-Omni supports simultaneous input of both speech and text. However, in voice assistant scenarios, additional text information may be necessary for optimal performance.

    \item Instruction following variability: Most models typically struggle to effectively follow instructions, particularly when dealing with tasks not involved in their training process. 

    \item Severe performance degradation: Introducing speech modalities into LLMs often leads to a marked decline in general knowledge, semantic understanding, and generation capabilities. A critical challenge is to preserve the performance of textual LLMs while supporting speech interactions.

    \item Unsatisfactory output speech quality: Some speech LLMs exhibit limited speech-text consistency and speech quality. These issues are not only related to the LLM used but also to the speech tokenization and training schemes.
\end{itemize}

\section{Conclusion}
In this paper, we introduce a comprehensive benchmark, SOVA-Bench, for evaluating generative speech LLMs from both semantic and acoustic perspectives. Evaluation experiments on novel speech-interactive models reveal variations in the quality of generated speech responses and a substantial degradation in performance compared to the cascade model, particularly in general knowledge and speech understanding. Considering the deficiencies in unified evaluation standards within the field of speech LLMs, especially voice assistants, SOVA-Bench provides a systematic framework for comparing model performance.


\section{Acknowledgements}
This work was supported by the Shanghai Jiao Tong University - Ant Group Intelligent Multimedia Joint Laboratory.

\bibliographystyle{IEEEtran}
\bibliography{mybib}

\begin{thebibliography}{10}
\providecommand{\url}[1]{#1}
\csname url@samestyle\endcsname
\providecommand{\newblock}{\relax}
\providecommand{\bibinfo}[2]{#2}
\providecommand{\BIBentrySTDinterwordspacing}{\spaceskip=0pt\relax}
\providecommand{\BIBentryALTinterwordstretchfactor}{4}
\providecommand{\BIBentryALTinterwordspacing}{\spaceskip=\fontdimen2\font plus
\BIBentryALTinterwordstretchfactor\fontdimen3\font minus \fontdimen4\font\relax}
\providecommand{\BIBforeignlanguage}[2]{{%
\expandafter\ifx\csname l@#1\endcsname\relax
\typeout{** WARNING: IEEEtran.bst: No hyphenation pattern has been}%
\typeout{** loaded for the language `#1'. Using the pattern for}%
\typeout{** the default language instead.}%
\else
\language=\csname l@#1\endcsname
\fi
#2}}
\providecommand{\BIBdecl}{\relax}
\BIBdecl

\bibitem{hurst2024gpt}
A.~Hurst, A.~Lerer, A.~P. Goucher, A.~Perelman, A.~Ramesh, A.~Clark, A.~Ostrow, A.~Welihinda, A.~Hayes, A.~Radford \emph{et~al.}, ``{GPT-4o} system card,'' \emph{arXiv preprint arXiv:2410.21276}, 2024.

\bibitem{xie2024mini}
Z.~Xie and C.~Wu, ``{Mini-Omni: Language models can hear, talk while thinking in streaming},'' \emph{arXiv preprint arXiv:2408.16725}, 2024.

\bibitem{fang2024llama}
Q.~Fang, S.~Guo, Y.~Zhou, Z.~Ma, S.~Zhang, and Y.~Feng, ``{LLaMA-Omni: Seamless speech interaction with large language models},'' \emph{arXiv preprint arXiv:2409.06666}, 2024.

\bibitem{defossez2024moshi}
A.~D{\'e}fossez, L.~Mazar{\'e}, M.~Orsini, A.~Royer, P.~P{\'e}rez, H.~J{\'e}gou, E.~Grave, and N.~Zeghidour, ``Moshi: a speech-text foundation model for real-time dialogue,'' \emph{arXiv preprint arXiv:2410.00037}, 2024.

\bibitem{huang2024dynamic}
C.-y. Huang, K.-H. Lu, S.-H. Wang, C.-Y. Hsiao, C.-Y. Kuan, H.~Wu, S.~Arora, K.-W. Chang, J.~Shi, Y.~Peng \emph{et~al.}, ``{Dynamic-SUPERB}: Towards a dynamic, collaborative, and comprehensive instruction-tuning benchmark for speech,'' in \emph{ICASSP 2024-2024 IEEE International Conference on Acoustics, Speech and Signal Processing (ICASSP)}.\hskip 1em plus 0.5em minus 0.4em\relax IEEE, 2024, pp. 12\,136--12\,140.

\bibitem{huang2024dynamic2}
C.-y. Huang, W.-C. Chen, S.-w. Yang, A.~T. Liu, C.-A. Li, Y.-X. Lin, W.-C. Tseng, A.~Diwan, Y.-J. Shih, J.~Shi \emph{et~al.}, ``{Dynamic-SUPERB Phase-2}: A collaboratively expanding benchmark for measuring the capabilities of spoken language models with 180 tasks,'' \emph{arXiv preprint arXiv:2411.05361}, 2024.

\bibitem{wang2024audiobench}
B.~Wang, X.~Zou, G.~Lin, S.~Sun, Z.~Liu, W.~Zhang, Z.~Liu, A.~Aw, and N.~F. Chen, ``{AudioBench}: A universal benchmark for audio large language models,'' \emph{arXiv preprint arXiv:2406.16020}, 2024.

\bibitem{yang2024air}
Q.~Yang, J.~Xu, W.~Liu, Y.~Chu, Z.~Jiang, X.~Zhou, Y.~Leng, Y.~Lv, Z.~Zhao, C.~Zhou \emph{et~al.}, ``{AIR-Bench}: Benchmarking large audio-language models via generative comprehension,'' \emph{arXiv preprint arXiv:2402.07729}, 2024.

\bibitem{chen2024voicebench}
Y.~Chen, X.~Yue, C.~Zhang, X.~Gao, R.~T. Tan, and H.~Li, ``{VoiceBench}: Benchmarking llm-based voice assistants,'' \emph{arXiv preprint arXiv:2410.17196}, 2024.

\bibitem{tangsalmonn}
C.~Tang, W.~Yu, G.~Sun, X.~Chen, T.~Tan, W.~Li, L.~Lu, Z.~Ma, and C.~Zhang, ``{SALMONN:Towards generic hearing abilities for large language models},'' in \emph{The Twelfth International Conference on Learning Representations,{ICLR} 2024, Vienna, Austria, May 7-11, 2024}.

\bibitem{zhang2023speechgpt}
D.~Zhang, S.~Li, X.~Zhang, J.~Zhan, P.~Wang, Y.~Zhou, and X.~Qiu, ``{SpeechGPT}: Empowering large language models with intrinsic cross-modal conversational abilities,'' in \emph{Findings of the Association for Computational Linguistics: EMNLP 2023}, 2023, pp. 15\,757--15\,773.

\bibitem{chu2024qwen2}
Y.~Chu, J.~Xu, Q.~Yang, H.~Wei, X.~Wei, Z.~Guo, Y.~Leng, Y.~Lv, J.~He, J.~Lin \emph{et~al.}, ``Qwen2-audio technical report,'' \emph{arXiv preprint arXiv:2407.10759}, 2024.

\bibitem{siuzdak2024snac}
H.~Siuzdak, F.~Gr{\"o}tschla, and L.~A. Lanzend{\"o}rfer, ``{SNAC}: Multi-scale neural audio codec,'' in \emph{Audio Imagination: NeurIPS 2024 Workshop AI-Driven Speech, Music, and Sound Generation}.

\bibitem{hsu2021hubert}
W.-N. Hsu, B.~Bolte, Y.-H.~H. Tsai, K.~Lakhotia, R.~Salakhutdinov, and A.~Mohamed, ``Hubert: Self-supervised speech representation learning by masked prediction of hidden units,'' \emph{IEEE/ACM transactions on audio, speech, and language processing}, vol.~29, pp. 3451--3460, 2021.

\bibitem{kong2020hifi}
J.~Kong, J.~Kim, and J.~Bae, ``{HiFi-GAN}: Generative adversarial networks for efficient and high fidelity speech synthesis,'' \emph{Advances in neural information processing systems}, vol.~33, pp. 17\,022--17\,033, 2020.

\bibitem{polyak2021speech}
A.~Polyak, Y.~Adi, J.~Copet, E.~Kharitonov, K.~Lakhotia, W.-N. Hsu, A.~Mohamed, and E.~Dupoux, ``Speech resynthesis from discrete disentangled self-supervised representations,'' in \emph{INTERSPEECH 2021-Annual Conference of the International Speech Communication Association}, 2021.

\bibitem{WestlakeOmni}
Xinchen-AI, ``{Westlake-Omni},'' \url{https://github.com/xinchen-ai/Westlake-Omni}, 2024.

\bibitem{wang2024freeze}
X.~Wang, Y.~Li, C.~Fu, L.~Xie, K.~Li, X.~Sun, and L.~Ma, ``{Freeze-Omni}: A smart and low latency speech-to-speech dialogue model with frozen llm,'' \emph{arXiv preprint arXiv:2411.00774}, 2024.

\bibitem{veluri2024beyond}
B.~Veluri, B.~Peloquin, B.~Yu, H.~Gong, and S.~Gollakota, ``Beyond turn-based interfaces: Synchronous llms as full-duplex dialogue agents,'' in \emph{Proceedings of the 2024 Conference on Empirical Methods in Natural Language Processing}, 2024, pp. 21\,390--21\,402.

\bibitem{zhang2024omniflatten}
Q.~Zhang, L.~Cheng, C.~Deng, Q.~Chen, W.~Wang, S.~Zheng, J.~Liu, H.~Yu, and C.~Tan, ``{OmniFlatten}: An end-to-end {GPT} model for seamless voice conversation,'' \emph{arXiv preprint arXiv:2410.17799}, 2024.

\bibitem{li2025baichuan}
Y.~Li, J.~Liu, T.~Zhang, S.~Chen, T.~Li, Z.~Li, L.~Liu, L.~Ming, G.~Dong, D.~Pan \emph{et~al.}, ``{Baichuan-Omni-1.5} technical report,'' \emph{arXiv preprint arXiv:2501.15368}, 2025.

\bibitem{joshi2017triviaqa}
M.~Joshi, E.~Choi, D.~S. Weld, and L.~Zettlemoyer, ``{TriviaQA}: A large scale distantly supervised challenge dataset for reading comprehension,'' in \emph{Proceedings of the 55th Annual Meeting of the Association for Computational Linguistics (Volume 1: Long Papers)}, 2017, pp. 1601--1611.

\bibitem{panayotov2015librispeech}
V.~Panayotov, G.~Chen, D.~Povey, and S.~Khudanpur, ``Librispeech: an asr corpus based on public domain audio books,'' in \emph{2015 IEEE international conference on acoustics, speech and signal processing (ICASSP)}.\hskip 1em plus 0.5em minus 0.4em\relax IEEE, 2015, pp. 5206--5210.

\bibitem{zhao2024librisqa}
Z.~Zhao, Y.~Jiang, H.~Liu, Y.~Wang, and Y.~Wang, ``{LibriSQA}: A novel dataset and framework for spoken question answering with large language models,'' \emph{IEEE Transactions on Artificial Intelligence}, 2024.

\bibitem{lee18d_interspeech}
C.-H. Lee, S.-L. Wu, C.-L. Liu, and H.~yi~Lee, ``{Spoken SQuAD}: A study of mitigating the impact of speech recognition errors on listening comprehension,'' in \emph{Interspeech 2018}, 2018, pp. 3459--3463.

\bibitem{busso2008iemocap}
C.~Busso, M.~Bulut, C.-C. Lee, A.~Kazemzadeh, E.~Mower, S.~Kim, J.~N. Chang, S.~Lee, and S.~S. Narayanan, ``{IEMOCAP}: Interactive emotional dyadic motion capture database,'' \emph{Language resources and evaluation}, vol.~42, pp. 335--359, 2008.

\bibitem{ardila2020common}
R.~Ardila, M.~Branson, K.~Davis, M.~Kohler, J.~Meyer, M.~Henretty, R.~Morais, L.~Saunders, F.~Tyers, and G.~Weber, ``{Common Voice}: A massively-multilingual speech corpus,'' in \emph{Proceedings of the Twelfth Language Resources and Evaluation Conference}, 2020, pp. 4218--4222.

\bibitem{taori2023stanford}
R.~Taori, I.~Gulrajani, T.~Zhang, Y.~Dubois, X.~Li, C.~Guestrin, P.~Liang, and T.~B. Hashimoto, ``Stanford alpaca: an instruction-following llama model (2023),'' \emph{URL https://github. com/tatsu-lab/stanford\_alpaca}, vol.~1, no.~9, 2023.

\bibitem{baba2024t05}
K.~Baba, W.~Nakata, Y.~Saito, and H.~Saruwatari, ``The t05 system for the voicemos challenge 2024: Transfer learning from deep image classifier to naturalness mos prediction of high-quality synthetic speech,'' \emph{arXiv preprint arXiv:2409.09305}, 2024.

\bibitem{du2024cosyvoice}
Z.~Du, Q.~Chen, S.~Zhang, K.~Hu, H.~Lu, Y.~Yang, H.~Hu, S.~Zheng, Y.~Gu, Z.~Ma \emph{et~al.}, ``Cosyvoice: A scalable multilingual zero-shot text-to-speech synthesizer based on supervised semantic tokens,'' \emph{arXiv preprint arXiv:2407.05407}, 2024.

\bibitem{radford2023robust}
A.~Radford, J.~W. Kim, T.~Xu, G.~Brockman, C.~McLeavey, and I.~Sutskever, ``Robust speech recognition via large-scale weak supervision,'' in \emph{International conference on machine learning}.\hskip 1em plus 0.5em minus 0.4em\relax PMLR, 2023, pp. 28\,492--28\,518.

\bibitem{mini2}
Z.~Xie and C.~Wu, ``{Mini-Omni2}: Towards open-source {GPT-4o} with vision, speech and duplex capabilities,'' \emph{arXiv preprint arXiv:2410.11190}, 2024.

\bibitem{zeng2024glm}
A.~Zeng, Z.~Du, M.~Liu, K.~Wang, S.~Jiang, L.~Zhao, Y.~Dong, and J.~Tang, ``{GLM-4-Voice}: Towards intelligent and human-like end-to-end spoken chatbot,'' \emph{arXiv preprint arXiv:2412.02612}, 2024.

\end{thebibliography}

\end{document}